# STEMPix: A Phase-Transition-Material-Based Pixel Sensor for Resolving Edge-Movement Direction

Md Rahatul Islam Udoy[1], Sumeet Kumar Gupta[2], Deep Jariwala[3], and Ahmedullah Aziz[1*]
[1]Department of Electrical Engineering and Computer Science, University of Tennessee, Knoxville, TN 37996, USA
[2]Elmore Family School of Electrical and Computer Engineering, Purdue University, West Lafayette, Indiana 47906, USA
[3]Electrical & Systems Engineering, University of Pennsylvania, Philadelphia, PA 19104, USA
[*]Email: aziz@utk.edu.

***Abstract*— This paper proposes a spatio-temporal edge-movement direction pixel (STEMPix) for generating compact direction-aware edge movement information inside a CMOS-compatible image sensor array. The proposed design targets specialized sensing applications where local boundary movement is more important than full-frame intensity reconstruction. Instead of transferring full multi-bit frames for external processing, STEMPix generates a 3-bit local edge direction code (LEDC) by combining pixel-level temporal change information with neighboring-pixel spatial edge information. We design the architecture using a two-tier organization, where the photodiode layer is separated from the computation layer to preserve light-collection area while accommodating the additional in-array processing circuitry. The proposed circuit is evaluated through HSPICE transient simulations. The estimated implementation achieves a horizontal pitch of 1.73 $\mu$m, a vertical pitch of 2.36 $\mu$m, and a geometric fill factor of 95.47%. The average active switching energy is 0.465 fJ per LEDC operation across representative edge-movement cases. The proposed STEMPix operation also supports global-shutter capture and dynamic thresholding. These results indicate that STEMPix can provide a compact and scalable front-end representation for edge-movement-aware sensing systems.**

## I. Introduction

Detecting the direction of edge movement is useful in applications where the important information is not the full image intensity, but how object boundaries change over time [1], [2]. For example, an expanding fire front, a growing crack, a moving blast boundary, a spreading biological colony, or a changing object contour can often be described by the local movement of its edges [3], [4]. Similar information can also support autonomous vehicles, defense monitoring, high-speed scene understanding, structural health monitoring, wildfire observation, and scientific imaging of dynamic processes [5]–[7]. In these applications, a compact representation of local edge movement can be more useful than repeatedly transferring and processing full multi-bit image frames.

Conventional image sensors usually capture pixel intensity values and send the data to peripheral circuits or external processors for feature extraction [8], [9]. This approach is flexible, but it creates a sensor-to-processor bottleneck when the pixel count or frame rate increases [10], [11]. The readout of full multi-bit frames also increases data bandwidth, power consumption, and latency [12]. Moreover, analog pixel values are sensitive to noise and variations before digitization [13]. These issues motivate in-sensor and in-pixel processing approaches, where useful visual features are extracted closer to the sensing location [14]. However, in-pixel processing should not be understood as simply moving conventional processor circuits into every pixel. Such a direct mapping would significantly increase pixel pitch, reduce spatial resolution, and increase power consumption. Instead, useful and compact operations must be carefully selected, and the pixel array must be designed so that the required computation can be accommodated efficiently. For edge-movement direction detection, this means that the sensor should not need to reconstruct or process a full image. It should only extract the local temporal and spatial information needed to determine how a binary edge moves.

Several prior works have explored on-chip edge extraction and event-based sensing. Edge-detection CMOS image sensors have demonstrated edge-image generation using column or readout peripheral circuits [15], [16]. These works retain simple conventional pixels, but the edge information is extracted after the pixel signals reach the readout path. Event-based image sensors generate temporal activity information closer to the pixel level [17], [18], but their outputs represent temporal events rather than local edge-movement direction. Earlier focal-plane visual motion measurement sensors also explored motion extraction near the sensor plane, including event-driven, and retinomorphic architectures for in-sensor motion detection and motion-direction processing [19]–[23]. These works demonstrate growing interest in extracting dynamic visual information directly within or near the sensing array. However, such architectures can become difficult to scale to larger arrays because of interconnection requirements and large motion-detection circuits. Therefore, a scalable pixel-array architecture that locally combines temporal changes with spatial edge information to explicitly encode edge-movement direction remains desirable.

Stacked image sensors provide an important path for integrating additional computation without sacrificing the photosensitive area [24], [25]. In a stacked organization, the sensing layer can be separated from the computation layer, and hybrid bonding can provide dense vertical connections between tiers. Such stacked CMOS image sensor organizations with hybrid bonding have been experimentally demonstrated [26], [27]. This integration direction is attractive for the proposed architecture because the photodiode can remain in the sensing layer while the event-generation and edge-computation circuits can be placed in another layer.

In this work, we present a circuit- and array-level design study of a spatio-temporal edge-movement direction pixel

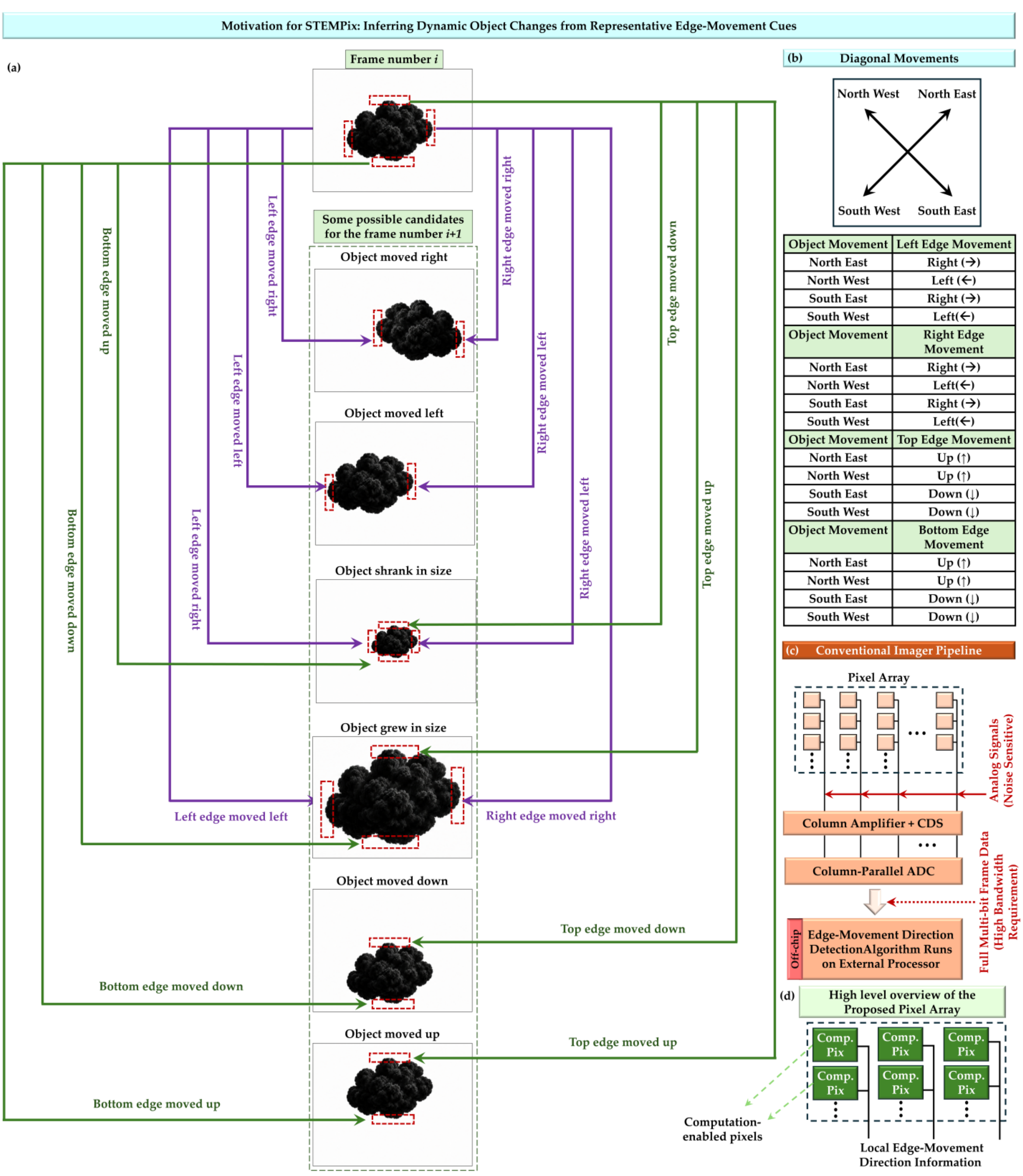

| Object Movement | Left Edge Movement |
|---|---|
| North East | Right (→) |
| North West | Left (←) |
| South East | Right (→) |
| South West | Left(←) |
| **Object Movement** | **Right Edge Movement** |
| North East | Right (→) |
| North West | Left(←) |
| South East | Right (→) |
| South West | Left(←) |
| **Object Movement** | **Top Edge Movement** |
| North East | Up (↑) |
| North West | Up (↑) |
| South East | Down (↓) |
| South West | Down (↓) |
| **Object Movement** | **Bottom Edge Movement** |
| North East | Up (↑) |
| North West | Up (↑) |
| South East | Down (↓) |
| South West | Down (↓) |

**Fig. 1:** Motivation of Spatio-Temporal Edge-Movement Direction Pixel (STEMPix) design. **(a)** Representative images of an abstract object in consecutive frames ($i$ and $i+1$) and the corresponding edge movements for horizontal translation, vertical translation, shrinking, and expansion. These examples show that object-level changes can be estimated from the direction of local edge movement. The examples show a dark object on a bright background, and the same principle can be applied to the reverse scenario, i.e., a bright object on a dark background. **(b)** Extension of the edge-movement interpretation to diagonal object movement, and the table shows corresponding orthogonal movements. **(c)** Conventional image sensing and processing pipeline, where full multi-bit frame data are read out and processed externally, which creates a bottleneck. **(d)** Proposed computation-enabled pixel array, where local edge-movement direction information is generated inside the array and can be read out at the column end.

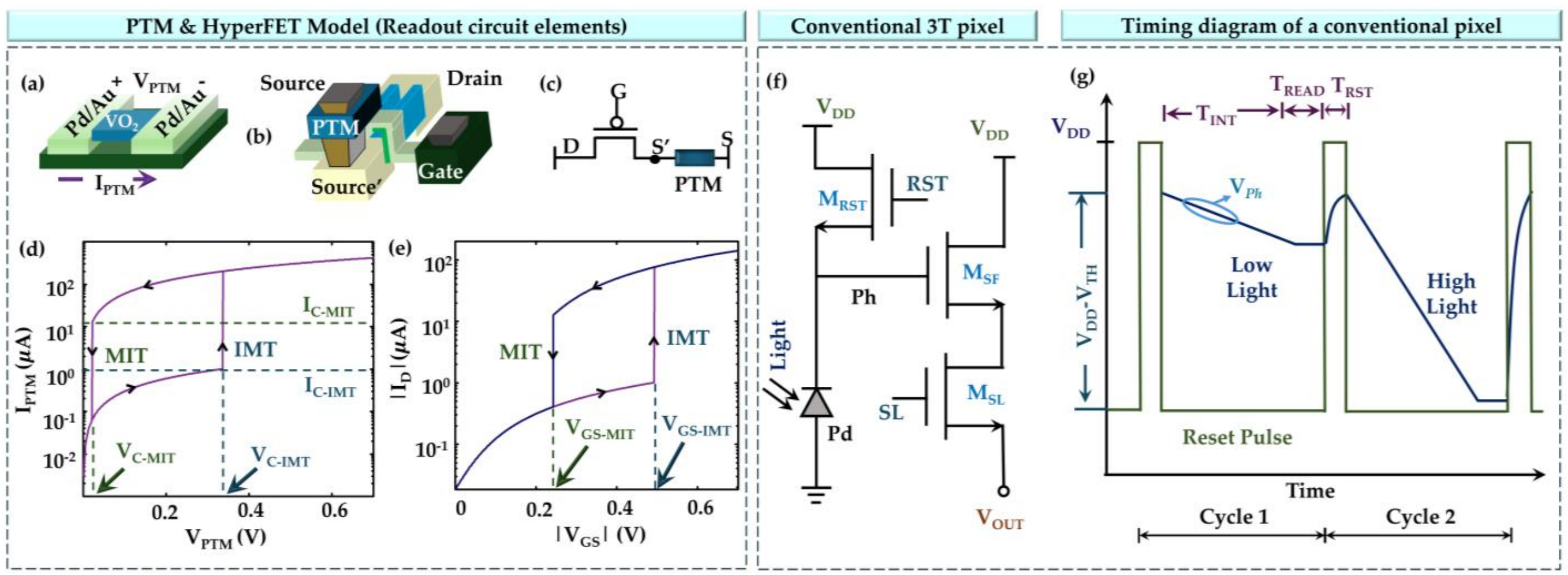

**Fig. 2:** **(a)** Phase Transition Material (PTM) structure. Here, Pd/Au is the metal electrode, and the blue region is the $VO_2$ channel. **(b)** HyperFET structure, which is constructed by integrating the PTM in the source terminal of a MOSFET. **(c)** A P-type HyperFET circuit symbol. **(d)** Current vs. voltage characteristics of a PTM. The device goes through insulator to metal transition (IMT) at a critical voltage $V_{C\text{-}IMT}$ and corresponding current $I_{C\text{-}IMT}$ . On the other hand, MIT stands for metal to insulator transition. **(e)** $|I_D|$ vs. $|V_{GS}|$ of the HyperFET. Here, $I_D$ is the drain current, and $V_{GS}$ is the gate to source voltage of the HyperFET. $V_{GS\text{-}IMT}$ is the threshold voltage for IMT switching. **(f)** Schematic of a conventional 3-transistor (3-T) pixel circuit. Here, $M_{RST}$, $M_{SF}$, and $M_{SL}$ are reset, source-follower, and pixel selector transistors, respectively. **(g)** Timing diagram of the 3-T pixel circuit. $T_{INT}$ is the light integration phase, $T_{READ}$ is the readout phase, and $T_{RST}$ is the reset phase. Higher illumination discharges the photodiode node 'Ph' faster during integration. The 1st cycle shows low level illumination and the 2nd cycle shows higher illumination.

(STEMPix) for local edge-movement direction detection inside the pixel array. We evaluate the proposed operation through HSPICE transient simulations and layout-based footprint estimation. The proposed architecture combines temporal event information from individual pixels with spatial binary-edge information between neighboring pixels to generate a compact 3-bit local edge direction code (LEDC). The circuit uses CMOS-compatible components and is designed toward monolithic integration, which is important for scalable array implementation without relying on nonstandard external processing hardware.

The proposed STEMPix is intended for specialized sensing applications where local edge-movement direction is more important than full-scene intensity reconstruction. Therefore, it is not meant to replace general-purpose cameras. Instead, the simulation results suggest that STEMPix can provide a compact front-end representation for dynamic scenes where edge movement itself carries the task-relevant information. The proposed timing scheme is also designed to support global-shutter-style operation. This is important for event and motion-related sensing because all pixels should make their binary decisions for the same time instant before sequential readout begins [28], [29]. Otherwise, motion-induced timing differences can distort the extracted event or edge-movement information [25], [30].

The rest of this paper is organized as follows. Section II presents the design motivation and technical background. Section III introduces the proposed STEMPix architecture, local edge direction coding, circuit configuration, and working mechanism. Section IV describes the array-level design and operation. Section V analyzes the footprint, active switching energy, and frame time. Section VI verifies dynamic thresholding and global-shutter operation. Section VII benchmarks the proposed architecture against representative edge-detection and event-based image sensors. Section VIII concludes the paper.

## II. Design Motivation and Technical Background

Object-level motion can be estimated from how its local edges change between consecutive frames. When an object changes position or size between two consecutive frames, the movement can be inferred from the local movement of its edges (Fig. 1(a)). For example, when the object moves to the right, the left and right object boundaries also move to the right. When the object shrinks, the left and right boundaries move toward each other, while the top and bottom boundaries also move inward. Similarly, when the object expands, the boundary movements indicate outward motion. Therefore, instead of processing the full image externally, we target local edge-movement cues as compact indicators of object-level change. Although Fig. 1(a) illustrates a dark object on a bright background, the same principle applies to the reverse case, where a bright object moves on a dark background. The same idea can also be extended to diagonal motion, as shown in Fig. 1(b). A diagonal object movement produces a corresponding combination of horizontal and vertical edge movements. For example, a northeast movement can be decomposed into rightward and upward local edge movements. Therefore, if the pixel array can locally detect whether an edge moved left, right, up, or down, then these local decisions can provide useful information about more complex object movement. This motivates the proposed design, where each local neighboring-pixel boundary produces a compact edge-direction code instead of requiring full-frame post-processing.

Fig. 1(c) shows the conventional imager pipeline. In a standard architecture, the pixel array produces analog pixel signals, which are read through column circuits such as column amplifiers, correlated double sampling (CDS), and column-parallel analog-to-digital converters (ADCs) [12], [31]. The resulting full multi-bit frame data are then processed externally to determine higher-level information such as edge movement. This pipeline is general and flexible, but it requires high data bandwidth because the full frame must be read out before edge-movement analysis. In contrast, Fig. 1(d) shows the high-level concept of the proposed computation-enabled pixel array. Here, the pixels locally generate edge-movement direction information inside the array. Therefore, the output becomes a compact local representation of motion-related changes rather than only raw frame data.

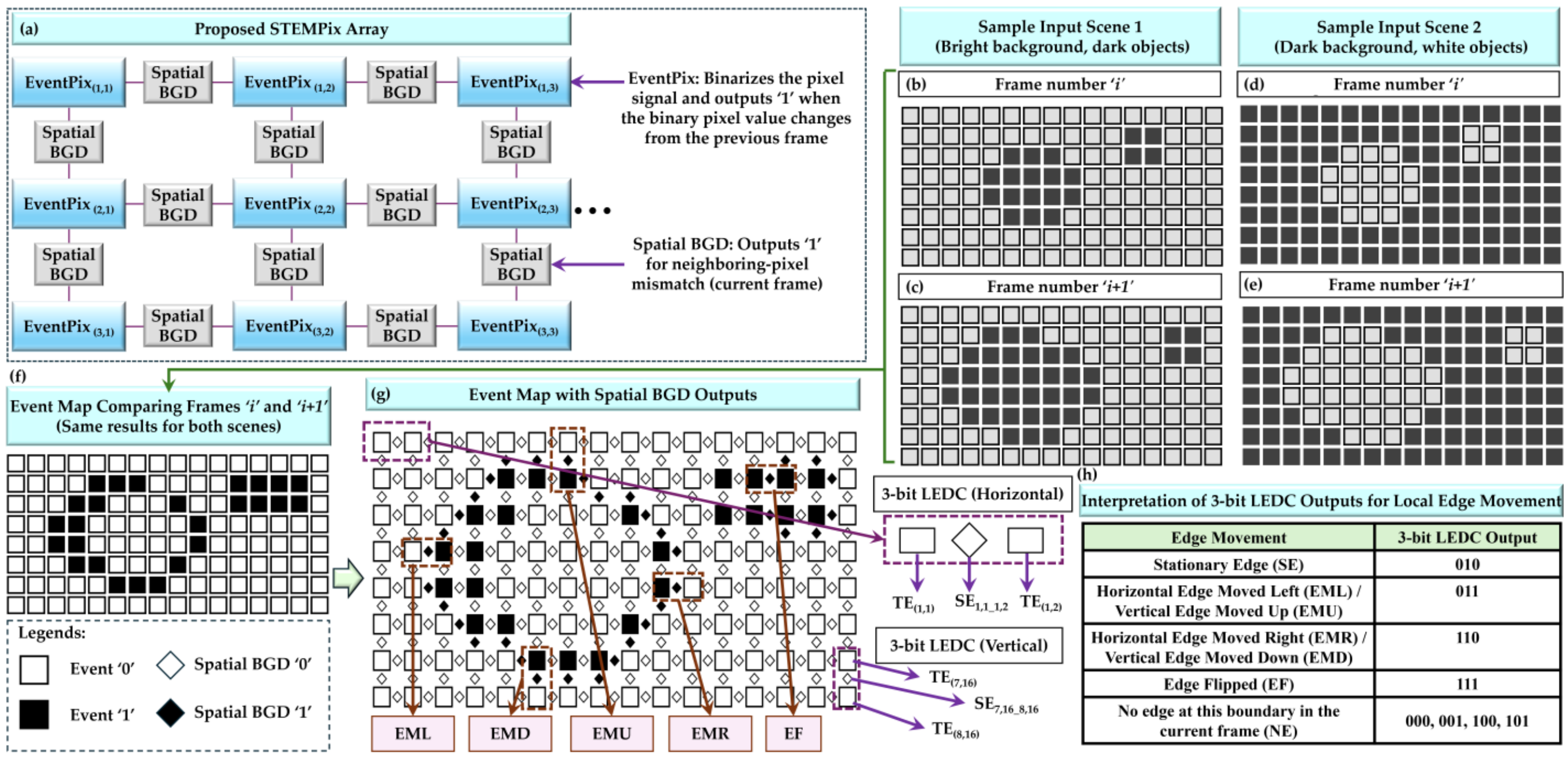

**Fig. 3:** Proposed spatio-temporal edge-movement direction pixel (STEMPix) array and local edge direction coding (LEDC) concept. **(a)** Proposed STEMPix array consisting of EventPix cells and spatial binary gradient detector (BGD) blocks. Each EventPix cell binarizes the pixel signal and generates a temporal event when the binary pixel value changes from the previous frame, while each spatial BGD block detects a mismatch between neighboring pixels in the current frame. **(b), (c)** Sample input scene with a bright background and dark objects in two consecutive frames. **(d), (e) S**ample input scene with a dark background and bright objects in two consecutive frames. **(f)** Temporal event map obtained by comparing frames $i$ and $i+1$. The same event map is produced for both sample scenes because the object movement is the same. **(g)** Combined event map with spatial BGD outputs used to generate LEDC codes throughout the array. Each 3-bit LEDC code is formed locally around a neighboring-pixel boundary. For a horizontal boundary, the left bit is the temporal event (TE) of the left pixel, the middle bit is the spatial edge (SE) between the two pixels, and the right bit is the TE of the right pixel. For a vertical boundary, the left, middle, and right bits correspond to the TE of the top pixel, the SE between the top and bottom pixels, and the TE of the bottom pixel, respectively. Therefore, overlapping LEDC codes are generated across the array for adjacent horizontal and vertical pixel pairs. Here (1,1) represents pixel position in the array and 1,1_1,2 means the spatial BGD in between pixel position (1,1) and (1,2). Examples indicate edge moved left (EML), edge moved down (EMD), edge moved up (EMU), edge moved right (EMR), and edge flipped (EF). **(h)** Interpretation of 3-bit LEDC outputs for horizontal and vertical local edge movement.

To understand the circuit-level baseline, Fig. 2(f) shows a conventional three-transistor (3T) active pixel sensor, which is the basic building block of an image sensor array [32], [33]. The circuit consists of a reset transistor $M_{RST}$, a source-follower transistor $M_{SF}$, and a selector transistor $M_{SL}$. During reset, $M_{RST}$ charges the photodiode node $Ph$ to approximately $V_{DD}$ - $V_{TH}$ if an NMOS reset transistor is used, where $V_{TH}$ is the threshold voltage of the reset transistor. If a PMOS reset transistor is used, $Ph$ can be reset to $V_{DD}$. During light integration, the photodiode discharges the $Ph$ node depending on the incident light intensity. A higher illumination level discharges $Ph$ faster, while a lower illumination level discharges it more slowly. During readout, $M_{SF}$ buffers the photodiode voltage, and $M_{SL}$ selects the pixel output. Fig. 2(g) shows this timing behavior for two cycles, where the second cycle has stronger illumination and therefore a faster drop in $V_{Ph}$.

Our proposed system builds on this light-dependent photodiode discharge behavior but changes how the pixel information is represented. In the proposed system, thresholding is a crucial step because it converts the light-dependent photodiode response into a binary pixel decision. For thresholding, we use a phase-transition material (PTM)-based HyperFET. Fig. 2(a) shows the PTM structure, where the $VO_2$ channel is contacted by Pd/Au electrodes. Recent studies have demonstrated $VO_2$ device fabrication using CMOS-compatible and BEOL-oriented integration processes, supporting its potential use as a compact thresholding element in integrated sensor circuits [34], [35]. Fig. 2(d) shows the current-voltage characteristic of the PTM. As the applied voltage increases, the device undergoes an insulator-to-metal transition (IMT) at the critical voltage $V_{C\text{-}IMT}$ and corresponding current $I_{C\text{-}IMT}$. When the voltage is reduced, the device returns through the metal-to-insulator transition (MIT). This abrupt transition provides a sharp switching behavior that is useful for threshold-based pixel binarization. At elevated temperatures, the critical voltage and current of the PTM may shift. For applications targeting a specific high-temperature operating range, these critical parameters can be engineered during the design phase. In our previous work, we demonstrated that the PTM switching threshold can be tuned by tailoring the device geometry [36].

Fig. 2(b) shows the HyperFET structure, where the PTM is integrated at the source terminal of a MOSFET. This structure makes the HyperFET compatible with CMOS circuit design because the PTM acts as an integrated source-side switching element while the main transistor remains a MOSFET. Numerous circuits have been demonstrated using the HyperFET as a crucial device for steep switching and thresholding applications [36]–[39]. Fig. 2(c) shows the corresponding p-type HyperFET circuit symbol. The PTM transition modulates the effective source behavior of the MOSFET, producing a steep switching response. Fig. 2(e) shows the $|I_D|$ vs. $|V_{GS}|$ characteristic of the HyperFET, where $V_{GS\text{-}IMT}$ indicates the gate-to-source voltage condition associated with IMT switching.

## III. STEMPix Architecture and Operation

We present the proposed spatio-temporal edge-movement direction pixel (STEMPix) architecture and its circuit-level implementation in this section.

### *A. Block Level Overview*

We first describe STEMPix from a high-level digital electronics perspective. In this work, an event refers to a change in the binary value of a pixel between two consecutive frames [40], [41]. If the current binary pixel value differs from the previous binary pixel value, the pixel generates a temporal event. If the two values are the same, no temporal event is generated. Therefore, the event output represents whether a local pixel-level change has occurred over time. We also define a binary local edge using the current binary values of neighboring pixels. If two adjacent pixels have different binary values, the boundary between them is treated as a local edge. If the two adjacent pixels have the same binary value, no local edge exists at that boundary. This binary edge representation does not require multi-bit intensity information. Instead, it uses the mismatch between neighboring binary pixel values to indicate the presence of a local spatial transition.

Fig. 3(a) shows the proposed STEMPix array structure. The array consists of EventPix cells and spatial binary gradient detector (BGD) blocks. Each EventPix cell generates a binary pixel value and compares the current binary value with the stored value from the previous frame to produce a temporal event output. The spatial BGD blocks compare the current binary values of neighboring EventPix cells to detect binary local edges. In this way, each EventPix cell provides temporal change information, while each BGD block provides spatial edge information between adjacent pixels.

When an object moves between two consecutive frames, only the pixels near the changing object boundary switch their binary values. This behavior is illustrated in Fig. 3(b) and Fig. 3(c) for a sample scene with a bright background and dark objects. The reverse contrast condition, where bright objects move on a dark background, is shown in Fig. 3(d) and Fig. 3(e). Although the object and background polarities are reversed, the movement pattern is the same. The temporal event map captures where local binary pixel values change between frames. As shown in Fig. 3(f), the same event map is obtained for both contrast cases because the event is defined by a change in binary pixel value, not by whether the object is brighter or darker than the background. Therefore, the temporal event output captures the location of frame-to-frame changes while remaining insensitive to the sign of the object-background contrast.

The local edge-direction representation requires both temporal change information and current-frame spatial edge information. In Fig. 3(g), the temporal event map is combined with the spatial binary gradient detector (BGD) outputs. The temporal event map identifies where pixel values changed between consecutive frames, while the spatial BGD outputs identify where local binary edges exist in the current frame. Together, these two types of binary information create a local representation around each neighboring-pixel boundary. For each adjacent pixel pair, the corresponding temporal event outputs and the spatial edge output form the structural basis for local edge-direction coding.

### *B. LEDC Generation and Interpretation*

We define the local edge direction code (LEDC) as a 3-bit code that describes the movement direction of a local binary edge between two neighboring pixels. The code is generated around a pixel-pair boundary by combining temporal change information from the two pixels with the spatial edge information between them. Therefore, each LEDC does not represent the state of a single pixel only. Instead, it represents the local spatio-temporal behavior of an edge around a neighboring-pixel pair. For a horizontal neighboring-pixel pair, the LEDC is formed as [$TE_L$, SE, $TE_R$]. Here, $TE_L$ is the temporal event output of the left pixel, $TE_R$ is the temporal event output of the right pixel, and SE is the spatial edge output between the two pixels in the current frame. $TE_L$ becomes 1 when the left pixel changes its binary value between two consecutive frames, and $TE_R$ becomes 1 when the right pixel changes its binary value between two consecutive frames. SE becomes 1 when the current binary values of the left and right pixels are different. Therefore, the left and right bits describe where a temporal change occurred, while the center bit describes whether a binary edge exists at the current boundary.

For a vertical neighboring-pixel pair, we use the same 3-bit structure, but the physical orientation changes. The LEDC is formed as [$TE_T$, SE, $TE_B$], where $TE_T$ is the temporal event output of the top pixel, $TE_B$ is the temporal event output of the bottom pixel, and SE is the spatial edge output between the top and bottom pixels. In this case, the left, center, and right bit names are code-position names rather than physical left, center, and right locations. Physically, the first and third bits correspond to the top and bottom pixels, respectively. This orientation convention allows the same 3-bit LEDC format to represent both horizontal and vertical local edge movement.

The LEDC interpretation is based on the local relationship between the two temporal event bits and the spatial edge bit. For a horizontal boundary, the code 011 means that the right pixel changed and a spatial edge exists between the two pixels in the current frame. This corresponds to an edge moving left. The code 110 means that the left pixel changed and a spatial edge exists, corresponding to an edge moving right. The code 010 means that a spatial edge exists, but neither neighboring pixel changed; therefore, the edge is stationary. The code 111 means that both neighboring pixels changed while a spatial edge exists, indicating an edge-flipped condition. The same logic applies to vertical boundaries, where 011 and 110 correspond to upward or downward edge movement depending on the selected top-to-bottom code orientation. This coding method is shown in Fig. 3(g) and Fig. 3(h). Fig. 3(g) shows that LEDC codes are generated locally throughout the array around adjacent horizontal and vertical pixel pairs. Because neighboring pairs overlap, one pixel can participate in multiple LEDC codes. For example, a pixel can contribute to a horizontal LEDC with its left or right neighbor and to a vertical LEDC with its top or bottom neighbor. This overlapping local coding is useful because edge movement is not assigned to an isolated pixel. Instead, it is assigned to the boundary where the binary spatial transition exists.

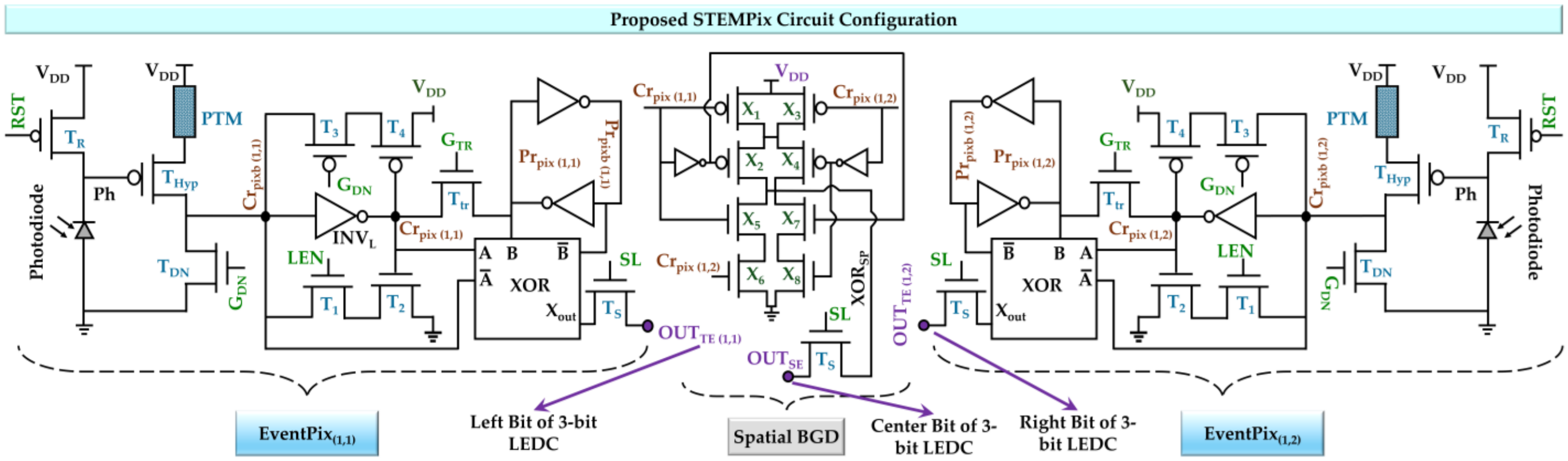

**Fig. 4:** Proposed STEMPix circuit configuration for two adjacent pixels and the corresponding binary gradient detector (BGD). Here, (1,1) and (1,2) denote pixel coordinate positions. In each EventPix cell, $T_R$ is the reset transistor, and $T_{Hyp}$ together with the PTM forms the HyperFET for thresholding. $T_{DN}$ is the discharge-enable transistor. $T_1$, $T_2$, $T_3$, $T_4$, and $INV_L$ latch the HyperFET thresholding decision. The transfer transistor Ttr transfers the current pixel binary value $Cr_{pix}$ to the pixel storage node at the end of each cycle. The XOR gate compares the previous pixel binary value $Pr_{pix}$ with the current pixel binary value $Cr_{pix}$ to generate the temporal event output $OUT_{TE}$. *RST*, $G_{DN}$, *LEN*, $G_{TR}$, *and SL* are control signals representing reset, discharge enable, latch enable, pixel-value transfer enable, and selector signals, respectively. Transistors $X_1$-$X_8$ form the XOR circuit in the spatial BGD to generate the spatial edge output $OUT_{SE}$. The selector transistors $T_S$ select the temporal event and spatial edge outputs. The left, center, and right bits of the 3-bit LEDC (local edge direction coding) are obtained from $OUT_{TE(1,1)}$, $OUT_{SE}$, *and* $OUT_{TE(1,2)}$, respectively.

The 3-bit LEDC is a compact representation of local edge movement. To determine the movement direction of a local binary edge, at least three pieces of binary information are needed: whether the first pixel changed, whether a spatial edge exists between the two pixels in the current frame, and whether the second pixel changed. If only the two temporal event bits are used, the code can show which pixels changed, but it cannot confirm whether a current-frame edge exists at the boundary. If only the spatial edge bit and one temporal event bit are used, the code cannot distinguish which side of the boundary changed. Therefore, the 3-bit LEDC provides the minimum binary code length needed to encode local edge-movement direction using this neighboring-pixel formulation. By generating LEDC locally, the proposed STEMPix array produces a direction-aware representation directly at the pixel-array level. The output is not a full multi-bit image frame and not only a binary event map. Instead, each LEDC combines temporal event detection and spatial edge detection into a compact code that indicates the local direction or state of edge movement.

### C. *Circuit Configuration*

We implement STEMPix by combining an event-generating pixel cell with a neighboring-pixel spatial comparison block. Fig. 4 shows the proposed circuit configuration for two adjacent STEMPix cells and the corresponding spatial binary gradient detector (BGD). Each STEMPix cell, denoted as EventPix in Fig. 4, contains four main parts: the reset path, the HyperFET-based thresholding path, the latch and storage nodes, and the temporal-event generation block. The two adjacent cells are labeled by their pixel coordinates, (1,1) and (1,2), to show how neighboring pixels are connected for local edge-direction coding. In each EventPix cell, $T_R$ is the reset transistor connected to the photodiode node. $T_{Hyp}$ and the phase-transition material (PTM) form the HyperFET used for thresholding. $T_{DN}$ connects the thresholding path to the discharge-control signal $G_{DN}$. The latch and storage portion of each EventPix cell stores the binary pixel states needed for temporal comparison. $T_1$, $T_2$, $T_3$, $T_4$, and $INV_L$ form the latch structure. The current-cycle binary pixel value is represented by *Crpix*, while the previous-cycle binary pixel value is represented by *Prpix*. The transfer transistor $T_{tr}$ connects *Crpix* to the *Prpix* storage node under the $G_{TR}$ control signal. This structure allows each pixel to keep both the present and previous binary states locally.

The temporal-event generation block is implemented using an XOR gate inside each EventPix cell. This XOR gate takes *Crpix* and *Prpix* as inputs and produces *Xout*. The selector transistor $T_S$ connects *Xout* to the temporal-event output line $OUT_{TE}$ during readout. Therefore, the EventPix cell contains all local elements required to produce a binary pixel state and expose its temporal-event output to the array-level readout line. The neighboring-pixel spatial comparison block is labeled as the spatial BGD in Fig. 4. In the shown implementation, transistors X1-X8 form an XOR circuit that compares the current binary pixel values of the two adjacent EventPix cells, $Crpix_{(1,1)}$ and $Crpix_{(1,2)}$. The output of this block is $XOR_{SP}$, which is connected to the spatial-edge output line $OUT_{SE}$ through a selector transistor $T_S$. Structurally, this BGD block is placed between neighboring EventPix cells because it operates on a pixel-pair boundary rather than on a single pixel. We use the term BGD instead of only calling this block an XOR gate because the architectural role of the block is binary gradient detection. In this implementation, an XOR circuit is sufficient because a mismatch between two neighboring binary pixel values represents a binary local edge. However, the BGD terminology keeps the architecture general. Other BGD circuit implementations may also be used to perform the same neighboring-pixel comparison depending on design constraints such as area, speed, power, logic style, or technology node.

TABLE I: Phase Transition Material (PTM) parameters

| Parameter Name | Parameter Definition | Parameter Values |
|---|---|---|
| $L_{PTM}$, $A_{PTM}$ | Length, Area | 42nm, 1260 $nm^2$ |
| $J_{C\text{-}IMT}$ | Critical current density for IMT | $8\times10^4$ A/$cm^2$ |
| $\rho_M$ | Metallic state resistivity | $5\times10^{-3}$ Ω. cm |
| $J_{C\text{-}MIT}$ | Critical current density for MIT | $1\times10^6$ A/$cm^2$ |
| $\rho_{INS}$ | Insulating state resistivity | 1 Ω. cm |

The control signals in Fig. 4 define the external interface of the EventPix and BGD blocks. *RST*, $G_{DN}$, *LEN*, $G_{TR}$, and *SL* represent reset, discharge enable, latch enable, pixel-value transfer enable, and selector signals, respectively. The supply rails VDD and GND are shared by the circuit blocks. During readout, the selected temporal-event and spatial-edge signals are available as $OUT_{TE}$ and $OUT_{SE}$, respectively. For the horizontal neighboring-pixel pair shown in Fig. 4, the 3-bit local edge direction code (LEDC) is assembled from the outputs of these blocks. The left bit comes from $OUT_{TE(1,1)}$, the center bit comes from $OUT_{SE}$, and the right bit comes from $OUT_{TE(1,2)}$. Thus, the circuit configuration physically maps the LEDC structure to two adjacent EventPix cells and one boundary-level BGD block.

### *D. Circuit Working Mechanism*

We verify the working mechanism of the proposed STEMPix circuit using circuit simulation in HSPICE, an industry-grade circuit simulation tool [42]. We implement the transistor-level circuit using 14 nm FinFET technology and the PTM-MG transistor model for the FinFET devices [43]. The phase-transition material (PTM) model used in the HyperFET is calibrated to an experimentally demonstrated device [37], [44]–[46]. The main PTM-related parameters, including device dimensions and resistivity values in the insulating and metallic states, are listed in Table I.

As shown in Fig. 5, the operation of STEMPix follows five phases in each cycle: reset, light integration, thresholding, readout, and pixel-value storage. The corresponding control signals *RST, SL*, $G_{DN}$, *LEN*, and $G_{TR}$ are shown in Fig. 5(a)-(c). The notation (1,1) and (1,2) denotes the positions (row number, column number) of the two adjacent pixels used in this example. Since a p-type transistor is used for reset, RST = 0 enables the reset operation. During the light integration phase, the photodiode node voltage *Vph* decreases according to the incident light intensity, as shown in Fig. 5(d) and (e). A higher light intensity discharges *Vph* faster and can make *Vph* fall below $V_{REF}$ before the thresholding time. We encode this condition as *Crpix* = 0. A lower light intensity discharges *Vph* more slowly, so *Vph* remains above $V_{REF}$ at the thresholding time. We encode this condition as *Crpix* = 1. The thresholding phase converts the light-dependent photodiode response into the current binary pixel value. At the selected thresholding time $T_{th}$, $G_{DN}$ enables the discharge path and *LEN* enables the latch. The effective reference voltage $V_{REF}$ is set by the $V_{GS\text{-}IMT}$ condition of the HyperFET. If *Vph* falls below $V_{REF}$ at or before $T_{th}$, the internal node *Crpixb* becomes high and the current pixel binary value *Crpix* becomes low. If *Vph* remains above $V_{REF}$ at *Tth*, *Crpixb* remains low and *Crpix* becomes high. This thresholding behavior is shown through the *Vph, Crpixb, and Crpix* waveforms in Fig. 5(d)-(i).

The first cycle establishes the previous-pixel value used for temporal comparison in Cycle 2. In the first cycle, the temporal event readout is not valid because *Prpix* has not yet been established. After the first-cycle *Crpix* value is generated, $G_{TR}$ transfers *Crpix* to *Prpix* during the pixel-storage phase. In the next cycle, a new *Crpix* value is generated and compared with the stored *Prpix* value. The XOR output *Xout* becomes high when *Crpix* and *Prpix* do not match, indicating a temporal event, and remains low when the two values match. This temporal comparison and the corresponding temporal-event readout are shown in Fig. 5(j) and Fig. 5(k).

In parallel, the spatial BGD compares the current binary values of the two adjacent pixels. The spatial XOR output $XOR_{SP}$ becomes high when the two current *Crpix* values are different, indicating a binary local edge at the boundary between pixels (1,1) and (1,2). If the two current *Crpix* values are the same, $XOR_{SP}$ remains low. The spatial comparison and the selected spatial-edge readout are shown in Fig. 5(l) and (m). During readout, the LEDC for this horizontal pixel pair is formed as [$OUT_{TE(1,1)}$, $OUT_{SE}$, $OUT_{TE(1,2)}$].

The three example cases in Fig. 5 verify how the temporal-event and spatial-edge signals produce the final LEDC. In case 1 (Fig. 5(a)-(m)), pixel (1,1) experiences high light in both cycles, so its *Crpix* remains 0 and $OUT_{TE(1,1)}$ = 0. Pixel (1,2) changes from high light in cycle 1 to low light in cycle 2, so its *Crpix* changes from 0 to 1 and $OUT_{TE(1,2)}$ = 1. In cycle 2, the current *Crpix* values of pixels (1,1) and (1,2) are different, so $OUT_{SE}$ = 1. Therefore, the resulting LEDC is 011, indicating that the local edge moved left. In case 2 (Fig. 5(n)-(z)), pixel (1,1) changes from high light in cycle 1 to low light in cycle 2, so its *Crpix* changes from 0 to 1 and $OUT_{TE(1,1)}$ = 1. Pixel (1,2) changes from low light in cycle 1 to high light in cycle 2, so its *Crpix* changes from 1 to 0 and $OUT_{TE(1,2)}$ = 1. In cycle 2, the current *Crpix* values of the two pixels are still different, so $OUT_{SE}$ = 1. Therefore, the resulting LEDC is 111, indicating an edge-flipped condition. In case 3 (Fig. 5(aa)-(mm)), pixel (1,1) experiences high light in both cycles, so its *Crpix* remains 0 and $OUT_{TE(1,1)}$ = 0. Pixel (1,2) experiences low light in both cycles, so its *Crpix* remains 1 and $OUT_{TE(1,2)}$ = 0. Since the current *Crpix* values of the two pixels are different in cycle 2, $OUT_{SE}$ = 1. Therefore, the resulting LEDC is 010, indicating a stationary edge.

The same circuit operation applies to other local edge movements, such as edge moving right, edge moving up, and edge moving down. Only the relative pixel positions and the selected neighboring-pixel boundary change. For a horizontal boundary, the LEDC uses the left and right pixel temporal-event outputs with the spatial edge between them. For a vertical boundary, the LEDC uses the top and bottom pixel temporal-event outputs with the spatial edge between them. Thus, the working mechanism remains the same across horizontal and vertical edge-direction detection.

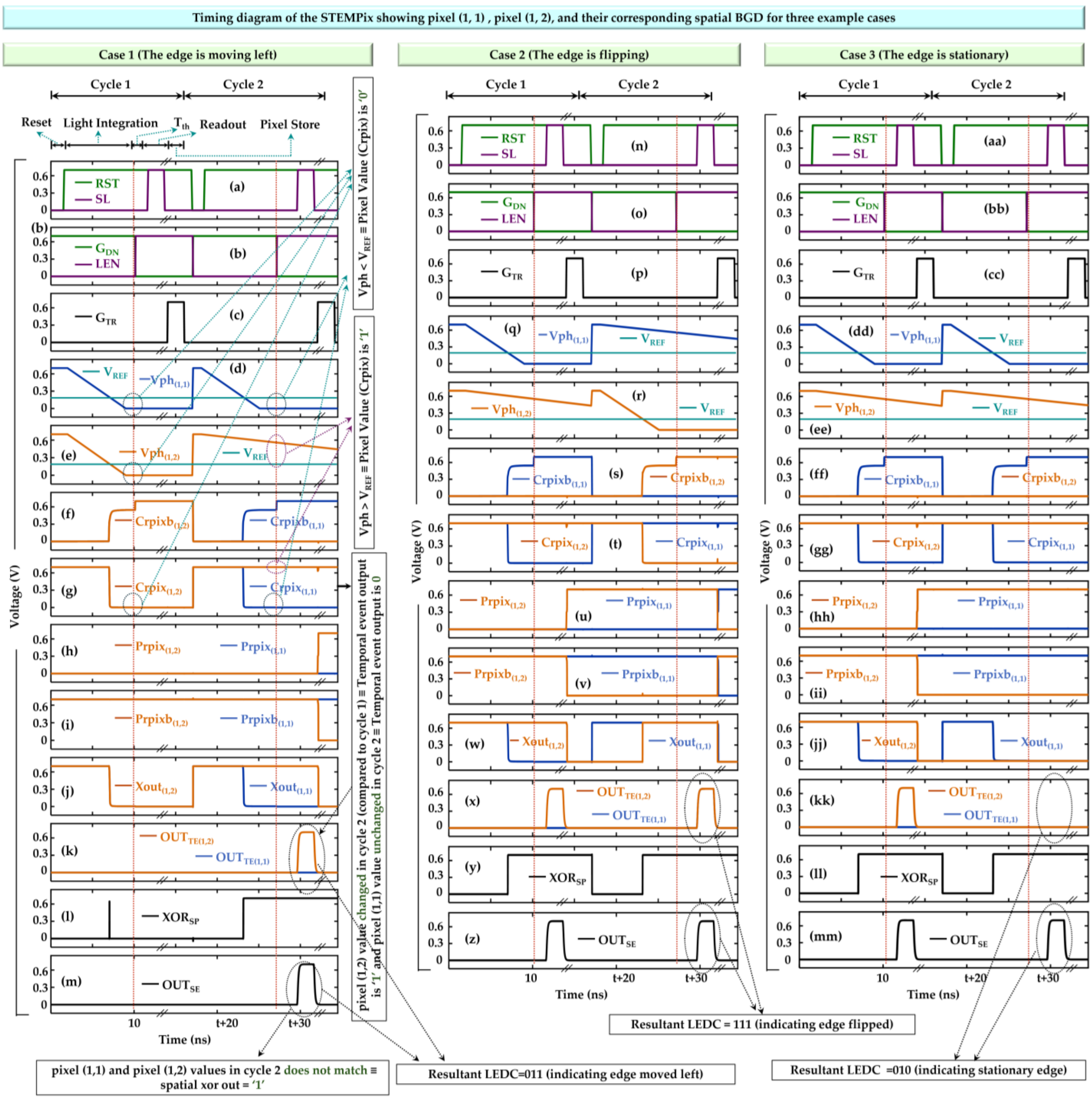

**Fig. 5:** Timing diagram of the proposed STEMPix circuit for two adjacent pixels and their corresponding spatial binary gradient detector (BGD). Three representative cases are shown: edge moved left, edge flipped, and stationary edge. **(a)-(c)** Control signals *RST*, *SL*, $G_{DN}$, LEN, and $G_{TR}$. Since a p-type (PMOS) transistor is used for reset, *RST* = 0 enables the reset operation. Each cycle consists of five phases: reset, light integration, thresholding (Tth), readout, and pixel storage. **(d), (e)** Photodiode node voltages *Vph* of the two adjacent pixels. A higher light intensity discharges *Vph* faster. If *Vph* falls below the reference voltage $V_{REF}$ at or before Tth, where $V_{REF}$ is set by the $V_{GS\text{-}IMT}$ of the HyperFET, the internal node *Crpixb* becomes high and the current pixel binary value *Crpix* becomes low. If *Vph* remains above $V_{REF}$ at or before Tth, *Crpixb* remains low and *Crpix* becomes high. Therefore, the pixel value is binarized using the HyperFET threshold, and *Crpix* stores the current-cycle binary pixel value. The first-cycle temporal event readout is not valid because the previous-cycle pixel value is not yet valid. Before the next cycle starts, *Crpix* is transferred to the previous-pixel storage node *Prpix*. In the next cycle, the new *Crpix* value is generated after reset, light integration, and thresholding. **(j)** XOR output *Xout*. For each pixel, *Xout* becomes high when *Crpix* and *Prpix* do not match, indicating a temporal event, and remains low otherwise. **(l)** Spatial XOR output $XOR_{SP}$, which becomes high when the *Crpix* values of adjacent pixels do not match, indicating a local binary edge. During readout, *SL* brings the temporal event output to $OUT_{TE}$ and the spatial edge output to $OUT_{SE}$, as shown in **(k)** and **(m)**. These outputs form the 3-bit LEDC. The resulting LEDC outputs are 011, 111, and 010 for cases 1, 2, and 3, indicating edge moved left, edge flipped, and stationary edge, respectively. Since readout is sequential and the timing diagram shows only two pixels, the time axis is broken during the readout phase. **(n)-(z)** show the waveforms for case 2, and **(aa)-(mm)** show the waveforms for case 3.

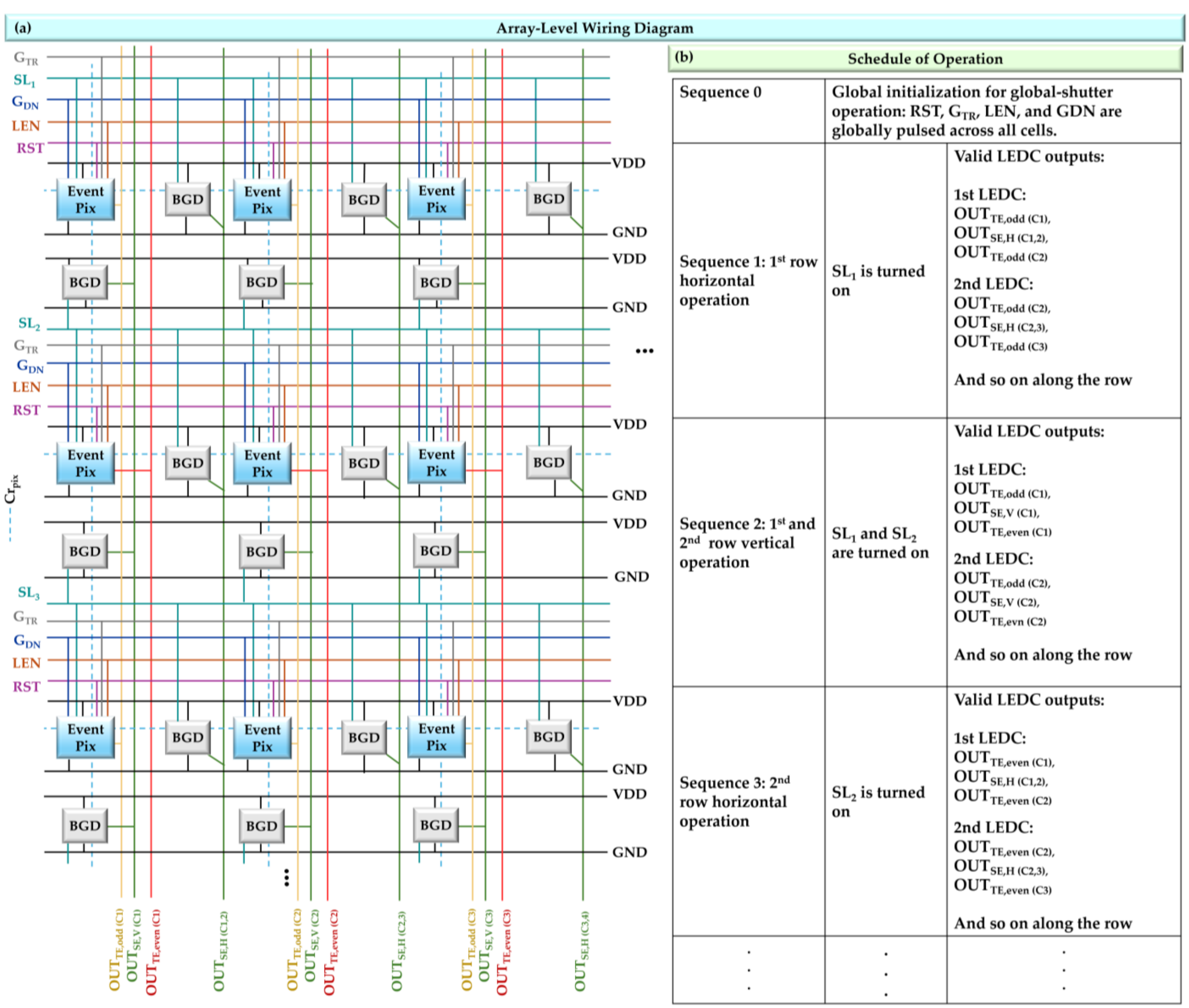

**Fig. 6:** Array-level wiring and operation schedule of the proposed STEMPix array. **(a)** Array-level wiring diagram showing EventPix cells and BGD blocks. $G_{TR}$, $SL$, $G_{DN}$, $LEN$, $RST$, VDD, and GND are shared along each row. The temporal event outputs from EventPix cells in odd-numbered rows are collected at the column end through $OUT_{TE,odd}$ lines, while those from even-numbered rows are collected through $OUT_{TE,even}$ lines. The spatial edge outputs from the BGDs located at horizontal boundaries between adjacent EventPix cells are collected through $OUT_{SE,H}$ lines, while those from the BGDs located at vertical boundaries are collected through $OUT_{SE,V}$ lines. C1, C2, C3, etc. denote the column numbers. **(b)** Schedule of operation for generating valid LEDC outputs. Only the readout phase is performed sequentially. The reset, light integration, thresholding, and pixel-storage phases are performed simultaneously across all pixels and BGDs in the array to support global-shutter operation.

## IV. Array-Level Design and Operation

We extend the STEMPix operation from a two-pixel example to an array-level structure by sharing the control signals across rows and collecting the temporal-event and spatial-edge outputs through dedicated readout lines. The array-level design must support two requirements. First, all pixels should capture and store their binary decisions at the same effective time to preserve a global-shutter-like operation. Second, the generated temporal-event and spatial-edge outputs should be read out in an ordered way so that valid local edge direction codes (LEDCs) can be reconstructed across the array.

The array-level wiring diagram is shown in Fig. 6(a). Each row contains EventPix cells, and the spatial binary gradient detector (BGD) blocks are placed at the boundaries between neighboring EventPix cells. Horizontal BGD blocks compare adjacent pixels within the same row, while vertical BGD blocks compare pixels between two neighboring rows. The control signals *RST, $G_{DN}$, LEN, GTR,* and *SL* are routed along the rows to coordinate reset, thresholding, latching, pixel-value storage, and readout. The supply rails VDD and GND are also shared across the row. The array generates LEDC outputs by selecting the appropriate row or row pair during readout. The operation schedule is shown in Fig. 6(b). Sequence 0 performs global initialization for global-shutter operation. In this sequence, *RST, $G_{TR}$, LEN, and $G_{DN}$* are globally pulsed across all cells. After this initialization, readout proceeds sequentially, while reset, light integration, thresholding, and pixel-value storage are performed simultaneously across the full array. Therefore, all pixels store their binary values for the same frame interval before row-wise readout begins.

For horizontal LEDC generation, only one row needs to be selected because the two pixels and the corresponding horizontal spatial BGD are located in the same row. For example, when $SL_1$ is turned on, the first horizontal LEDC in row 1 is formed using $OUT_{TE,odd(C1)}$, $OUT_{SE,H(C1,2)}$, and $OUT_{TE,odd(C2)}$, which correspond to the temporal event of the first pixel, the horizontal spatial edge between the first and second pixels, and the temporal event of the second pixel. The next LEDC along the same row uses $OUT_{TE,odd(C2)}$, $OUT_{SE,H(C2,3)}$, and $OUT_{TE,odd(C3)}$. In this way, overlapping horizontal LEDC codes are generated along the selected row. For vertical LEDC generation, two adjacent rows must be selected together because the two pixels forming the vertical boundary are located in different rows. For example, to generate vertical LEDCs between the first and second rows, $SL_1$ and $SL_2$ are turned on together. The temporal-event output from the first-row pixel appears on the $OUT_{TE,odd}$ line, while the temporal-event output from the second-row pixel appears on the $OUT_{TE,even}$ line. The vertical BGD output between these two pixels appears on the $OUT_{SE,V}$ line. These three outputs form the vertical LEDC for that column. The same procedure continues across the columns to generate vertical LEDC outputs along the row boundary. This vertical readout requirement motivates the separation of the temporal-event output lines according to row parity. Temporal-event outputs from odd-numbered rows are collected through $OUT_{TE,odd}$ lines, while temporal-event outputs from even-numbered rows are collected through $OUT_{TE,even}$ lines. If all rows shared the same temporal-event output line, the top and bottom pixels in a selected vertical pair would drive the same line simultaneously during vertical LEDC generation. By using separate odd-row and even-row temporal-event output lines, the array can read the top-pixel temporal event, the vertical spatial-edge output, and the bottom-pixel temporal event as three distinct LEDC bits.

The same readout pattern repeats for the rest of the array. After completing the first-row horizontal operation and the vertical operation between the first and second rows, the array proceeds to the second-row horizontal operation by turning on $SL_2$. Then, vertical operation between the second and third rows is performed by selecting $SL_2$ and $SL_3$ together. This alternating row-wise schedule allows the array to generate local edge-direction information for both horizontal and vertical neighboring-pixel boundaries.

We also estimate the physical layout of the proposed STEMPix implementation to evaluate the feasibility of placing the required circuit blocks within a pixel-array structure. The estimated second-layer layout is shown in Fig. 7. In the proposed two-layer implementation, the photodiode and reset transistor are placed in the first layer, while the second layer contains the thresholding, latching, storage, temporal-event generation, and spatial-edge generation circuits. This separation allows the photosensitive layer and the computation layer to be optimized with different layout constraints. The layout also reserves a 0.4 μm × 0.4 μm hybrid-bonding landing region to connect the photodiode layer to the second-layer circuit. This assumption is motivated by prior demonstrations of sub-micron-pitch in-pixel Cu-Cu bonding in stacked CMOS image sensors [26], [47] and 0.4 μm-pitch Cu-Cu hybrid bonding [27]. The landing region represents the physical interface between the sensing layer and the computation layer in the two-layer implementation.

## V. Footprint and Performance Analysis

### A. Footprint

We estimate the footprint of the proposed STEMPix using a two-tier implementation. In this organization, the first layer is used for light sensing, while the second layer is used for most of the computation circuitry. The first layer contains the photodiode, the reset transistor, and the hybrid-bonding landing region. The second layer contains the remaining STEMPix circuitry required for thresholding, latching, pixel-value storage, temporal-event generation, and spatial-edge generation, as illustrated in Fig. 7. This partitioning allows the computation circuitry to be moved away from the photosensitive layer while preserving most of the first-layer pixel footprint for light collection. The analysis estimates whether the additional circuits required for thresholding, temporal-event generation, and spatial-edge generation can be accommodated within a pixel-array-compatible pitch. The calculation is summarized in Table II.

The footprint estimation uses the pitch parameters of a 14 nm FinFET technology. Based on the reported technology scaling values in [48], we use a fin pitch (FP) of 42 nm, a gate pitch (GP) of 70 nm, and an interconnect metal pitch (MP) of 52 nm. From the estimated layout in Fig. 7, the horizontal pixel pitch is obtained by adding the reserved hybrid-bonding landing region, the EventPix length, and the spatial BGD length. This gives a horizontal pitch of 1730 nm. The vertical pixel pitch is obtained by adding the EventPix width and the spatial BGD width, resulting in a vertical pitch of 2360 nm (The hybrid-bonding landing-pad dimension is not added in this direction because the landing region is placed laterally beside the EventPix block in the estimated layout). Therefore, the total repeating pixel footprint is defined by the product of the horizontal and vertical pitches. The available active photodiode area is estimated from the first-layer footprint. Since the first layer contains the photodiode, the reset transistor, and the hybrid-bonding landing region, the reset-transistor area and the hybrid-bonding landing-pad area are subtracted from the total pixel footprint. This gives an estimation of the available active photodiode area listed in Table II. The subtraction provides a conservative estimate because the landing region is treated as unavailable for light collection. We report the geometric fill factor (FF) to quantify the fraction of the pixel footprint that remains available for the active photodiode region in the first layer. The geometric FF is calculated as the ratio between the available active photodiode area and the total pixel footprint. The proposed two-tier organization assumes backside illumination, where light enters through the backside of the photodiode tier; therefore, the computation tier and hybrid-bonding interconnects do not directly obstruct the optical path. Using the values in Table II, the estimated geometric FF is 95.47%. This is a layout-based estimate and does not include optical effects such as microlens focusing, isolation structures, or wavelength-dependent absorption. The high geometric FF is enabled by the two-tier organization, where the photodiode remains in the sensing layer

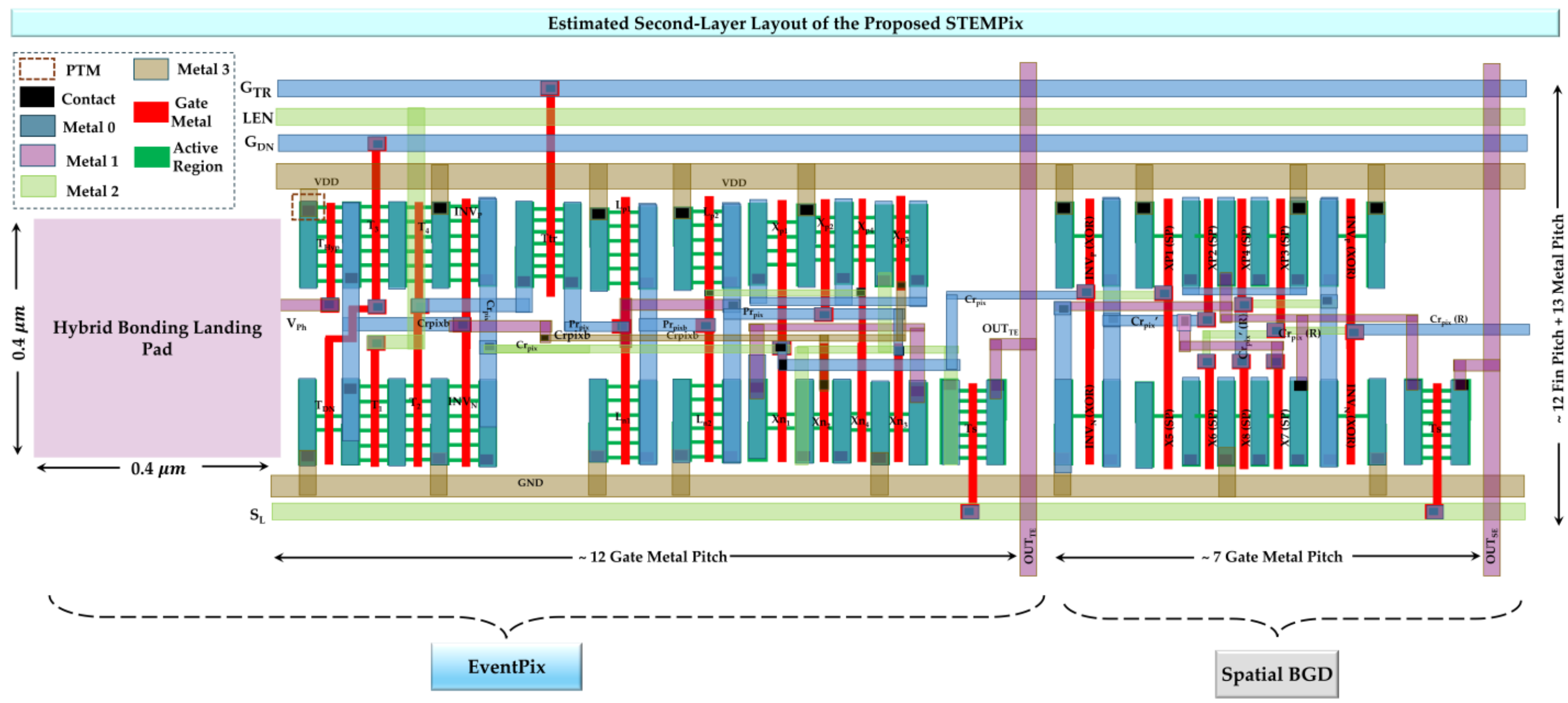

**Fig. 7:** Estimated second-layer layout of the proposed STEMPix (not drawn to scale). The layout shows the estimated placement of the EventPix circuit and the corresponding spatial BGD circuit in the second layer. In the proposed two-layer implementation, the first layer, not shown in this figure, contains the photodiode and the reset transistor. The second layer contains the remaining transistor circuits required for thresholding, latching, pixel-value storage, TE generation, and SE generation. The design uses 14 nm FinFETs. A hybrid-bonding landing pad is included to connect the photodiode layer with the second-layer circuit.

**TABLE II: Estimated Footprint Parameters of the Proposed STEMPix.**

| Parameters | Calculation | Value |
| --- | --- | --- |
| FinFET technology | - | 14 $nm$ |
| Number of metal layers | - | 4 |
| Eventpix length (horizontal) | $12 \times GP = 12 \times 70$ | 840 $nm$ |
| EventPix width (vertical) | $12 \times FP + 13 \times MP = 12 \times 42 + 13 \times 52$ | 1180 $nm$ |
| Spatial BGD length (horizontal) | $7 \times GP = 7 \times 70$ | 490 $nm$ |
| Spatial BGD width (vertical) | $Same\ as\ EventPix\ width$ | 1180 $nm$ |
| Horizontal pixel pitch | $Hybrid\ bonding\ landing\ pad + EventPix\ length + Spatial\ BGD\ length = 400 + 840 + 490$ | 1730 $nm$ |
| Vertical pixel pitch | $EventPix\ width + Spatial\ BGD\ width = 1180 + 1180$ | 2360 $nm$ |
| Available active photodiode area in the first layer | $2360 \times 1730 - reset\ transitor\ area - hybrid\ bonding\ landing\ pad$ | 3.898 $\mu m^2$ |
| Geometric fill factor (FF) | $\frac{photodiode\ area}{horizontal \times vertical\ pixel\ pitch}$ | 95.47% |

*GP=Gate Pitch, FP=Fin Pitch, MP=Interconnect Metal Pitch

while most of the computation circuitry is placed in the second layer.

## B. *Switching Energy Analysis*

We evaluate the active switching energy of the proposed STEMPix circuit using transient HSPICE simulation. The circuit activity depends on the local edge-movement condition because different LEDC outputs require different internal nodes to switch. Therefore, we estimate the energy for representative operation cases, including edge moving left, edge moving right, edge flipped, and stationary edge, and report the average value across these cases. For each operation case, we calculate the energy drawn from the relevant supply sources during the active switching intervals. If the $i$-th active interval has an average supply power of $P_{avg,i}$ and a duration of $\Delta t_i$, the energy contribution of that interval is $E_i = P_{avg,i} \times \Delta t_i$. The total active switching energy for one operation case is then obtained by summing the energy contributions from all selected active intervals and supply sources:

$$E_{case} = \sum_i P_{avg,i} \times \Delta t_i$$

Equivalently, this calculation represents the integration of supply power over the active switching windows:

$$E_{case} = \sum_i \int_{t_{i,start}}^{t_{i,end}} P_i(t)dt$$

Here, $P_i(t)$ is the instantaneous power drawn from the corresponding supply source during the $i$-th active interval. In the HSPICE measurement, we first identify the time windows in which the supply delivers switching current, calculate the average power in each window, multiply it by the corresponding time duration, and then sum all contributions. This avoids including long inactive intervals that do not contribute to active LEDC generation. Since different edge-movement cases activate different internal nodes, we calculate the active switching energy for multiple LEDC-generation cases and then average the results. The average active switching energy is calculated as

$$E_{avg} = \frac{1}{N}\sum_{k=1}^{N} E_{case,k}$$

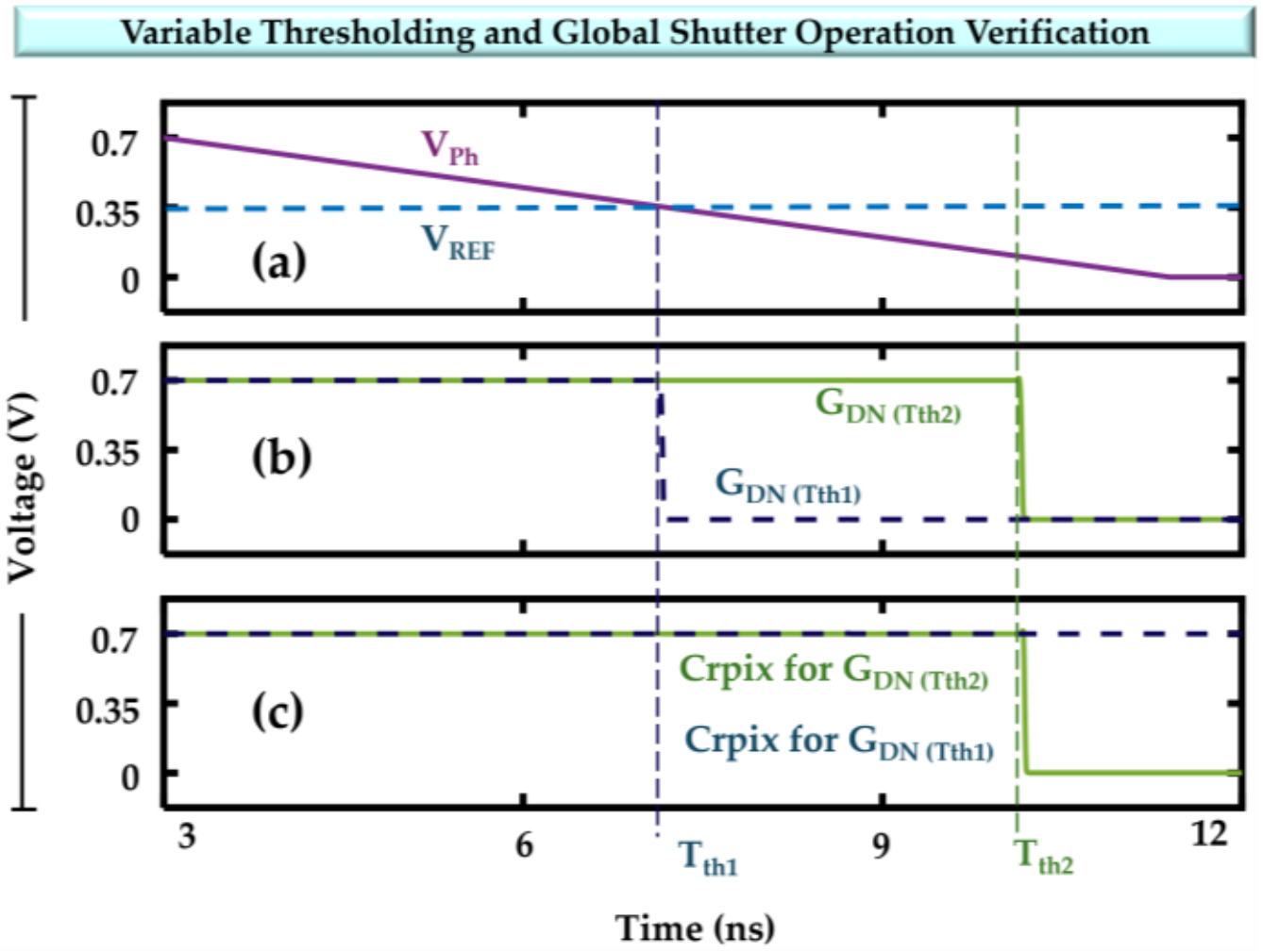

**Fig. 8:** Variable thresholding and global-shutter operation verification of the proposed STEMPix circuit. (a) Photodiode node voltage *Vph* and reference voltage $V_{REF}$. For the same illumination level, the final binary pixel value depends on the selected thresholding time. (b) Two example $G_{DN}$ signals are used to start thresholding (by making $G_{DN}$ low and *LEN* high) at different times, $T_{th1}$ and $T_{th2}$ (*LEN* is not shown in the figure). (c) Resulting *Crpix* values for the two thresholding times. At $T_{th1}$, *Vph* has not fallen below $V_{REF}$, so *Crpix* becomes high. At $T_{th2}$, *Vph* has already fallen below $V_{REF}$, so *Crpix* becomes low. This shows that the effective threshold can be controlled by changing the thresholding time. The figure also illustrates global-shutter operation verification: after the thresholding decision is latched, *Crpix* does not change even if *Vph* later falls below $V_{REF}$. Therefore, *Vph* must cross $V_{REF}$ before the selected thresholding time to affect the pixel value.

Where $N$ is the number of representative operation cases and $E_{case,k}$ is the active switching energy of the $k$-th case. Using this method, the average active switching energy of the proposed STEMPix operation is estimated as 0.465 fJ. This value represents the average active energy required to generate the local edge-direction output across the representative LEDC cases.

### C. Frame Time

We estimate the frame time from the array-level operation schedule. In the proposed array, reset, light integration, thresholding, and pixel-value storage are performed globally across all pixels. Therefore, these phases occur only once per frame before the sequential readout begins. From the simulated timing sequence, the time from reset to the start of the first row readout is 13.7 ns. After the global operation, the array reads the local edge direction code (LEDC) outputs row by row following the schedule in Fig. 6(b). For an array with $N_{row}$ rows, horizontal LEDC readout requires one row selection for each row. Therefore, the number of horizontal readout steps is $N_H = N_{row}$. Vertical LEDC readout requires selecting two adjacent rows together. Since the number of adjacent row pairs is $N_{row} - 1$, the number of vertical readout steps is $N_V = N_{row} - 1$. Thus, the total number of sequential readout steps required to complete one frame is $N_{read} = N_H + N_V = 2N_{row} - 1$. Each readout step is controlled by an SL pulse with a duration of 2 ns. Therefore, the total readout time is $T_{read} = (2N_{row} - 1) \times T_{SL}$. The total frame time is obtained by

$$T_{frame} = 13.7\,ns + (2N_{row} - 1) \times 2\,ns$$
$$= (4N_{row} + 11.7)\,ns$$

This expression shows that the frame time scales linearly with the number of rows.

## VI. Dynamic Thresholding and Global-Shutter Operation Verification

We verify two additional operating capabilities of the proposed STEMPix circuit: dynamic thresholding and global-shutter operation. Dynamic thresholding allows the effective binary decision boundary to be adjusted by changing the thresholding time, while global-shutter operation ensures that the stored binary pixel value does not change after the thresholding decision is latched.

The dynamic-thresholding concept relies on the time-dependent discharge of the photodiode node. Under a fixed illumination condition, the photodiode node voltage Vph decreases with time, as shown in Fig. 8(a). Therefore, even for the same light intensity, the binary pixel value can be changed by selecting a different thresholding time. If thresholding is performed earlier, Vph may still be above the reference voltage VREF. If thresholding is performed later, Vph may already have fallen below VREF. In this way, the selected thresholding time acts as an effective threshold-control parameter. In Fig. 8, we demonstrate this behavior using two thresholding times, $T_{th1}$ and $T_{th2}$. The thresholding operation is initiated by making $G_{DN}$ low and *LEN* high (*LEN* is not shown in the figure). The two $G_{DN}$ waveforms corresponding to $T_{th1}$ and $T_{th2}$ are shown in Fig. 8(b). At $T_{th1}$, *Vph* has not yet fallen below $V_{REF}$. Therefore, the pixel stores *Crpix* = 1. At $T_{th2}$, *Vph* has already fallen below $V_{REF}$. Therefore, the pixel stores *Crpix* = 0. The resulting *Crpix* waveforms are shown in Fig. 8(c). This result verifies that the proposed circuit can change the binary pixel decision by controlling the thresholding time, even when the illumination condition remains the same.

The same waveform also verifies the global-shutter behavior of the proposed operation. Once the thresholding decision is made and latched, *Crpix* remains fixed during the rest of the cycle. Therefore, a later crossing of *Vph* below $V_{REF}$ does not change the stored pixel value. This behavior is visible in the $T_{th1}$ case in Fig. 8(c), where *Crpix* remains high even though *Vph* falls below $V_{REF}$ after the thresholding instant. Thus, the pixel value is determined by whether *Vph* crosses $V_{REF}$ before the selected thresholding time, not by the final value of *Vph* at the end of the integration interval. This verification is important for array-level STEMPix operation. Since thresholding and latching can be applied globally across the array, all pixels can store their binary decisions at the same selected time. The stored *Crpix* values can then be used for temporal-event generation, spatial-edge generation, and sequential LEDC readout without being altered by additional photodiode discharge after the capture instant.

## VII. Benchmarking

Table III compares the proposed STEMPix with representative edge-extraction and event-generation image sensors. Since STEMPix generates local edge-movement direction information, while prior CMOS image sensor works mainly report either edge images or temporal events, the comparison is intended to position the proposed functionality relative to the closest available sensor architectures. The edge-extraction sensors in [15] and [16] use conventional 3T or 4T pixels, and the edge computation is performed in the column or readout peripheral circuitry. This keeps the pixel structure simple, but the extracted information becomes available only after the pixel signal reaches the readout path. In contrast, STEMPix moves the key operations closer to the sensing location by generating temporal event information inside the pixel and spatial edge information at the neighboring-pixel boundary. Therefore, the proposed architecture produces a direction-aware local code rather than only an edge image. The event-generation sensors in [17] and [18] perform temporal event detection closer to the pixel level. These works are more related to STEMPix in terms of local temporal activity sensing. However, their outputs are event streams or event frames, and additional processing is still needed to infer edge movement direction. STEMPix instead combines temporal event and spatial binary-edge information locally, so the output directly represents the local edge-movement state. The footprint and energy trends in Table III also follow the architectural differences. The prior edge-extraction works use larger technology nodes and include column/readout-level processing, while STEMPix uses an estimated 14 nm FinFET implementation and performs compact binary operations locally. Therefore, the reported STEMPix energy represents active switching energy for local LEDC generation. The comparison should be interpreted with this metric difference in mind. Still, the result indicates that the local binary operation needed for edge-movement direction detection can be implemented with very low switching energy.

TABLE III: Benchmarking

| Work | Operation | Process tech. | Transistors per pixel / cell | Processing location | Pixel pitch ($\mu m$) | Frame time / timing | Energy per pixel or event | Geometric FF | Stacked Sensor or not |
|---|---|---|---|---|---|---|---|---|---|
| This work | Edge extraction + detection of edge movement direction | 14 $nm$ | 23T per EventPix + 12T per spatial BGD | In-pixel and neighboring-pixel boundary | Horizontal= 1.73, Vertical= 2.36 | Tframe = 13.7 ns + (2Nrow - 1) × 2 ns | 0.465 fJ active switching energy per LEDC operation [Simulated] | 95.47% | Yes |
| [15] | Edge extraction | 0.18 $\mu m$ | 3T pixel | Column-end peripheral circuit | 8 | 240 fps, about 4.17 ms/frame | 721 pJ/pixel/frame [Derived from Measured Silicon] | - | No |
| [16] | Edge extraction | 0.18 $\mu m$ | 4T APS | Column-end peripheral circuit | 4.9 | 3200 fps for 160 × 120, about 312.5 us/frame; row readout 2.6 us | 70 pJ/pixel/frame [Measured (Silicon)] | 53% | No |
| [17] | event generation | 65 $nm$ | Several tens of transistors per pixel, exact count not reported | In-pixel event circuit | 4.95 | Event-driven, no conventional frame time; maximum event rate 1.3 Geps | 122 nW/pixel; total power 150 mW at high activity [Measured (Silicon)] | 22% | Yes |
| [18] | event generation | 65 $nm$ | Not reported | In-pixel event memory + group logic | 10 | 1 kefps event-frame operation, about 1 ms event-frame period; max event rate 180 Meps | 26 pJ/event dynamic energy; 18 nW/pixel static power [Derived from Measured Silicon] | 20% | No |

## VIII. Conclusion

This paper proposed STEMPix, a pixel sensor design for detecting edge-movement direction inside the image sensor array. By combining temporal pixel-change information with spatial binary-edge information, STEMPix generates a compact local edge direction code instead of relying only on full-frame intensity readout. The design study suggests that selected low-level visual processing can be moved closer to the sensing plane while preserving a compact pixel-array structure through a two-tier organization. The estimated implementation achieves a 1.73 $\mu$m × 2.36 $\mu$m pixel pitch, 95.47% geometric fill factor, and 0.465 fJ average active switching energy per LEDC operation. These results indicate that local edge-movement direction coding can be implemented with low circuit-level switching energy while maintaining most of the sensing-layer area for photodiode integration. Looking forward, STEMPix suggests a focused direction for pixel-array architectures that encode local motion-direction cues closer to the sensing plane. By combining compact device-level thresholding, local binary processing, and two-tier integration, this type of architecture may help future CMOS-compatible sensors generate edge-movement-aware front-end representations before full-frame readout. Such sensing approaches could support high-speed computer vision and scientific monitoring systems where local boundary movement direction is more important than full-scene reconstruction.